\documentclass[twocolumn,showpacs,amsmath,amssymb]{revtex4}

\usepackage{graphicx}
\usepackage{dcolumn}
\usepackage{bm}

\newcommand{\norm}[1]{\left\Vert#1\right\Vert}
\newcommand{\calH}{\mathcal{H}}
\newcommand{\calI}{\mathcal{I}}
\newcommand{\calL}{\mathcal{L}}
\newcommand{\up}{\uparrow}
\newcommand{\dn}{\downarrow}
\newcommand{\rme}{\mathrm{e}}
\newcommand{\rmi}{\mathrm{i}}
\newcommand{\Ne}{N_\mathrm{e}}
\newcommand{\Stot}{S_\mathrm{tot}}
\newcommand{\PhiG}{\Phi_\mathrm{G}}
\newcommand{\tg}[2]{{\tilde{g}(I_\uparrow^{#1};I_\downarrow^{#2})}}
\newcommand{\g}[2]{{{g}(I_\uparrow^{#1};I_\downarrow^{#2})}}
\newcommand{\La}{\Lambda}
\newcommand{\sumsigma}{\sum_{\sigma=\uparrow,\downarrow}}
\newcommand{\sumtwo}[2]%
{\mathop{\sum_{#1}}_{#2}}
\begin{document}

\title{Stability of ferromagnetism
in the Hubbard model on the kagom\'e lattice}

\author{Akinori Tanaka}
\author{Hiromitsu Ueda}%
\affiliation{%
Department of Applied Quantum Physics, Kyushu University, 
Fukuoka 812-8581, Japan
}%

\date{September 18, 2002}

\begin{abstract}
The Hubbard model on the kagom\'e lattice has 
highly degenerate ground states 
(the flat lowest band) in the corresponding
single-electron problem
and exhibits the so-called flat-band ferromagnetism in the many-electron 
ground states as was found by Mielke.
Here we study the model obtained by adding extra hopping terms 
to the above model.
The lowest single-electron band becomes dispersive, and there is 
no band gap between the lowest band and the other band. 
We prove that, at half-filling of the lowest band,  
the ground states of this perturbed model
remain saturated ferromagnetic
if the lowest band is nearly flat.
\end{abstract}

\maketitle
Numerous studies have been made on the Hubbard model, 
a tight-binding model of electrons with on-site interactions,
to understand
mechanisms for ferromagnetism in itinerant electron systems
in a simplified situation
~\cite{Lieb95,Hartmann95,Vollhardt99,Tasaki98}.
Recently, Mielke~\cite{Mielke} and Tasaki~\cite{Tasaki92} 
brought a significant breakthrough
in the field 
by proving 
that certain classes of
Hubbard models have ferromagnetic ground states.   
These models in common have multi single-electron bands 
containing dispersionless bands,  
and are called
flat-band Hubbard models.

Although these flat-band Hubbard models shed light on
the role of the Coulomb interaction in generating ferromagnetism,
the models with completely flat bands are still singular.
It is desirable to clarify
whether the flat-band ferromagnetism is stable against
perturbations which turn the flat bands into  dispersive bands.
As for Tasaki's version of flat-band Hubbard models,
local stability of the ferromagnetic ground state in perturbed
nearly-flat-band models was proved~\cite{Tasaki96}.
Tasaki also gave a concrete example of nearly-flat-band
Hubbard models in which he could prove that the ground
states are ferromagnetic~\cite{Tasaki95,Tasaki02}.
See \cite{Tanaka98,Sekizawa02} for related results.

As for Mielke's version of flat-band Hubbard models, on
the other hand, there have been no rigorous results about
stability (or instability) of ferromagnetism in
perturbed nearly-flat-band models~\cite{Tamura02}.
Here we note that 
there are essential differences between Mielke's and
Tasaki's models.
Mielke's models have simple structures where all the
lattice sites are identical, while Tasaki's models have
two different kinds of lattice sites.
Reflecting the lattice structures, there
are no band gaps in Mielke's models while there are finite
band gaps in Tasaki's models.
We stress that the problem of stability of ferromagnetism is
much more subtle and difficult in Mielke's models, where one might
encounter various low energy excitation modes which arise {from}
the gapless nature of the band structures.

In this Letter ,
we treat the model obtained by adding hopping terms 
to the Hubbard model on the kagom\'e lattice,
a typical example of Mielke's models~\cite{comment:other-line-graphs}.  
The added perturbation destroys flatness of the band,
but the band structure remains gapless.
We prove that our model has saturated ferromagnetic ground states
at half-filling of the lowest band, 
provided that the lowest band is nearly flat.
\begin{figure}[b]
\includegraphics[scale=1]{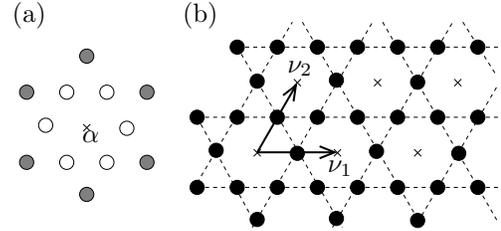}
\caption{\label{fig:lattice}
(a)~Local lattice $C_\alpha=C_\alpha^1\cup C_\alpha^2$. 
The open and gray circles represent sites in $C_\alpha^1$ and
$C_\alpha^2$, respectively. 
(b)~Lattice~$\La$.}
\end{figure}

\textit{Definition}.
We first define the reference 
triangular 
lattice $\calL$ as 
\begin{equation}
 \calL = \left\{ 
            n_1\nu_1 + n_2\nu_2 ~|~
            \mbox{\begin{minipage}{.2\textwidth}
                   $n_i\in\Bbb{Z}$ and $0\le n_i< L$ 
                   for $i=1,2$\end{minipage}}
         \right\}\,,
\end{equation}
where $\nu_1=(1,0)$, $\nu_2= (\frac{1}{2},\frac{\sqrt{3}}{2})$, and
$L$ is a positive integer. 
For each $\alpha\in\calL$ we define
\begin{eqnarray}
 & C_{\alpha}^1=\{x=n_1\frac{\nu_1}{2}+n_2\frac{\nu_2}{2}
                   ~|~\mbox{$n_1,n_2\in \Bbb{Z}$,$|x-\alpha|=\frac{1}{2}$} \},& \\
 & C_{\alpha}^2=\{x=n_1\frac{\nu_1}{2}+n_2\frac{\nu_2}{2}
                   ~|~\mbox{$n_1,n_2\in \Bbb{Z}$,$|x-\alpha|=\frac{\sqrt{3}}{2}$} \},& 
\end{eqnarray}
and ${C_\alpha=C_\alpha^1\cup C_\alpha^2}$~\cite{comment:norm}.
Then, the kagom\'e lattice $\La$ can be constructed 
as ${\La=\cup_{\alpha\in\calL}C_\alpha}$, 
where a site $x\in\La$ is generally counted four times in different
$C_\alpha$. (See Fig.\ref{fig:lattice}.)
Our lattice has open boundaries. 
Periodic lattices can be also treated with extra technical complication. 
We denote by $\Phi_0$ the state with no electrons and
denote by $c_{x,\sigma}$ and $c_{x,\sigma}^\dagger$ 
the annihilation and the creation 
operators, respectively, of an electron with spin~$\sigma$ 
at site~$x$ in~$\La$.
These operators satisfy the usual fermion anticommutation relations.
The number operator of an electron with spin $\sigma$ 
at site $x$ is defined as 
$n_{x,\sigma} = c_{x,\sigma}^\dagger c_{x,\sigma}$.
The total spin operators 
${\bm S}_\mathrm{tot}
  =(\Stot^{(1)},\Stot^{(2)},\Stot^{(3)})$ 
are defined as
$
 \Stot^{(i)} = \frac{1}{2}
              \sum_{x\in\La}
              \sum_{\sigma,\tau = \up,\dn}
              c_{x,\sigma}^\dagger
	      p_{\sigma\tau}^{(i)}
	      c_{x,\tau}
$
for $i=1,2,3$, where $p^{(i)}=[p^{(i)}]_{\sigma,\tau=\up,\dn}$
are the Pauli matrices. 
We denote by $\Stot(\Stot+1)$ the eigenvalue of 
$({\bm S}_\mathrm{tot})^2$.

Let us define our Hubbard Hamiltonian on $\La$.
First, for each $x\in C_\alpha^2$ we define fermion operator
$
 b_{(\alpha,x),\sigma} 
      = c_{x,\sigma} 
        + \sum_{y\in C_{\alpha}^1;|y-x|=\frac{1}{2}}c_{y,\sigma}\,.
$ 
We also define
$
 a_{\alpha,\sigma} = \sum_{y\in C_\alpha^1} \mu[\alpha,y]c_{y,\sigma} 
$ 
for $\alpha\in\calL$, where coefficients 
$\mu[\alpha,y]$ take either $+1$ or $-1$ 
and are chosen so that $\mu[\alpha,y]\mu[\alpha,y^\prime]=-1$ 
whenever $|y-y^\prime|=\frac{1}{2}$. 
To each $C_\alpha$ we associate the local Hamiltonian
\begin{eqnarray}
\calH_\alpha &=& -s\sumsigma a_{\alpha,\sigma}^\dagger a_{\alpha,\sigma}
                       +\frac{t}{3}\sumsigma
                       \sum_{x\in C_\alpha^2}
                       b_{(\alpha,x),\sigma}^\dagger b_{(\alpha,x),\sigma}\nonumber\\
              &&  + \frac{U}{4}\sum_{x\in C_\alpha}n_{x,\up}n_{x,\dn}\,,  
\end{eqnarray}
where $s$, $t$ and $U$ are positive parameters.  
Then, the Hubbard Hamiltonian on the whole lattice $\Lambda$ is defined  
as $\calH = \sum_{\alpha\in\calL} \calH_\alpha$.

\textit{Remarks}.
It is possible to rewrite $\calH$
into the standard form as 
${\calH = \sum_\sigma
        \sum_{x,y\in\La} t_{xy}c_{x,\sigma}^\dagger c_{y,\sigma}
         +\sum_{x\in\La} U_x n_{x,\up}n_{x,\dn}}$,
where the model parameters are given by
$U_x=U$, and 
$t_{xy}=2(t-s)$ if $x=y$,
$t_{xy}=t+s$ if $|x-y|=1/2$,
$t_{xy}=-s$ if $|x-y|=\sqrt{3}/2$,
$t_{xy}=s$ if $|x-y|=1$ and $x,y\in C_\alpha^1$ with some $\alpha$,
and $t_{xy}=0$ otherwise,
except for the sites close to the boundary.    
The single-electron dispersion relations 
(calculated in the model with periodic boundary conditions)
are given by
$E_1(k)=-2s(3-e(k))$, 
$E_2(k)=t(3-\sqrt{3+2e(k)})$, 
and 
${E_3(k)={t(3+\sqrt{3+2e(k)})}}$
with ${e(k)=\cos k_1 + \cos k_2 +\cos (k_1-k_2)}$, 
where ${k=k_1\nu_1^\ast+k_2\nu_2^\ast}$
is the wave vector expanded in terms of
reciprocal-lattice vectors ${\nu_1^\ast=(1,-\frac{1}{\sqrt{3}})}$ 
and ${\nu_2^\ast=(0,\frac{2}{\sqrt{3}})}$.
Note that $E_1(0)=E_2(0)=0$, which means that there is no gap 
between the lowest and the second lowest bands.

One readily finds that 
$\{a_{\alpha,\sigma}, b_{(\beta,x),\sigma}^\dagger\}=0$ 
for any $\alpha,\beta\in\calL$ and $x\in C_{\beta}^2$.
This implies that 
$\{a_{\alpha,\sigma}^\dagger\Phi_0\}_{\alpha\in\calL}$ 
spans the space corresponding to the lowest band.

If we set $s=0$, our model has highly degenerate single-electron
ground states, and becomes essentially the flat-band model of Mielke's
(although there is a difference in boundary conditions).
In this case, the model
exhibits flat-band ferromagnetism for all positive values of $U$.
In the model with $s>0$,
the situation is quite different
because double occupancies of lower energy states, 
which destroy the ferromagnetic order,
may reduce the total energy of the system.
It is indeed easy to prove 
that the ground states of our model 
has $\Stot=0$ (or $\frac{1}{2}$) for $U=0$, and cannot
exhibit saturated ferromagnetism
for sufficiently small $U$.
(See, for example, Sect.~3.3 of \cite{Tasaki98}).
The following theorem establishes that the ferromagnetic ground states
are stable for sufficiently large $t$ and $U$ 
when the electron number is $|\calL|$.

\textit{Main theorem}.
\textit{
Consider the Hubbard model defined as above
with the electron number $|\calL|$.
Then, there exist critical values $(t/s)_\mathrm{c}$ and 
$(U/s)_\mathrm{c}$, independent of the lattice size, 
such that, if both $t/s>(t/s)_\mathrm{c}$ and $U/s>(U/s)_\mathrm{c}$
are satisfied,
the ground states of the model have $\Stot = |\calL|/2$.
Furthermore the ground state is unique up to the degeneracy 
due to the rotational symmetry. 
}

In Tasaki's models, the stability of ferromagnetism may be, 
at least at a heuristic level, 
understood as a consequence of the band
gap separating the lowest nearly flat band {from} other bands.
The band gap enforces the electrons to occupy the lowest band
while the interaction rules out double occupancies of sites.
Then the situation is almost as in the flat-band models,
and the systems exhibit ferromagnetism.
To Mielke's models, which have no band gaps,
the above argument does not apply, and the origin
of the stability of ferromagnetism seems more subtle.
Nevertheless, our proof is based on essentially the same philosophy
as that of Tasaki's proof in \cite{Tasaki95}.
Namely, we first establish ferromagnetism
in a local model described by $\calH_\alpha$,
and then show 
that these local ferromagnetism can be ``connected'',
which results in macroscopic ferromagnetism in the whole system. 
The results of the analysis of $\calH_\alpha$
are summarized in the following lemma.

\textit{Lemma. If $t/s$ and $U/s$ are sufficiently large, 
the minimum eigenvalue of $\calH_\alpha$ is $-6s$ 
and any eigenstate $\Phi$ belonging to this eigenvalue 
is written as
\begin{equation}
\label{eq:condition1 in Lemma}
 \Phi=a_{\alpha,\up}^\dagger\Phi_{\up}
      +a_{\alpha,\dn}^\dagger\Phi_{\dn}\,,
\end{equation}
with appropriate 
states $\Phi_{\up}$ and $\Phi_{\dn}$.
Furthermore it satisfies 
\begin{equation}
\label{eq:condition2 in Lemma}
c_{x,\dn}c_{x,\up}\Phi = 0
\end{equation}
for all $x\in C_\alpha$.  
}

\textit{Proof of Lemma}.
Since all the local Hamiltonians are the translated copies 
of $\calH_0$, 
it suffices to prove the lemma for $\alpha=0$.
{From} now on, for convenience, 
we identify $C_0^1$ and $C_0^2$ with $\{0,2,\dots,10\}$
and $\{1,3,\dots,11\}$, respectively 
(we first label $(\frac{1}{2},0)$
as 0, then label the rest sites as
$1,\dots,11$ in the clockwise order).

We start by solving a single-electron problem for  
$\calH_0$.
Let ${I=\{0,\pm\frac{\pi}{3},\pm\frac{2\pi}{3},\pi\}}$. 
Then the eigenvalues are given by  
\begin{equation}
\label{eq:eigenvalue}
\varepsilon_1(p) = \left\{\begin{array}{ll}
                     0                   & \mbox{if $p\in I\backslash\{\pi\}$};\\
                     -6s                 & \mbox{if $p=\pi$}\,,
                   \end{array} \right.
\end{equation}
and $\varepsilon_2(p) =\frac{t}{3}(3+2\cos p)$ with $p\in I$. 
The eigenstate corresponding to $\varepsilon_1(p)$ 
is expressed as $d_{p,\sigma}^\dagger\Phi_0$ with 
\begin{equation}
 d_{p,\sigma} = \frac{1}{\sqrt{6(3+2\cos p)}}\sum_{l=0}^5\rme^{\rmi pl}(c_{2l,\sigma}
                                            -c_{2l-1,\sigma}-c_{2l+1,\sigma})
\end{equation}
(where $c_{-1,\sigma}$ is regarded as $c_{11,\sigma}$).
Note that the set
$\{d_{p,\sigma}^\dagger\Phi_0\}_{p\in I}$
is orthonormal since $\{d_{p,\sigma},d_{p^\prime,\sigma}^\dagger\}=\delta_{p,p^\prime}$.

We consider many-electron problem for $\calH_0$,
first in the limit $t,U\to\infty$.
Let $\Phi$ be a state on $C_0$ which has a finite energy in this limit.
Since all $\varepsilon_2(p)$ are infinite in the limit $t\to\infty$, 
$\Phi$ must be expanded as   
\begin{equation}
\label{eq:finite energy state}
\Phi = \sum_{I_\up,I_\dn\subset {I}}g(I_\up;I_\dn)\Phi(I_\up;I_\dn)
\end{equation}  
with complex coefficients $g(I_\up;I_\dn)$, where
\begin{equation}
\label{eq:basis Phi}
\Phi(I_\up;I_\dn) = \prod_{p\in I_\up}d_{p,\up}^\dagger
                    \prod_{p^\prime\in I_\dn}d_{p^\prime,\dn}^\dagger \Phi_0\,.
\end{equation}
Here, and throughout the present Letter, the products are ordered
in such a way that $d_{p,\up}^\dagger$(\textit{resp.} $d_{p,\dn}^\dagger$) 
is always on the left of
$d_{p^\prime,\up}^\dagger$(\textit{resp.} $d_{p^\prime,\dn}^\dagger$) if $p<p^\prime$. 
Since the on-site interaction 
$n_{x,\up}n_{x,\dn}=c_{x,\up}^\dagger c_{x,\dn}^\dagger
c_{x,\dn}c_{x,\up}$ 
is positive semidefinite,
the state $\Phi$ in the form of \eqref{eq:finite energy state}
must further satisfy
\begin{equation}
 \label{eq:ccPhi}
  \sum_{I_\up,I_\dn\subset {I}}
      g(I_\up;I_\dn)c_{x,\dn}c_{x,\up}\Phi(I_\up;I_\dn)=0
\end{equation}
for any $x\in C_0$ in order to have finite energy in the limit $U\to\infty$.
{From} \eqref{eq:eigenvalue}, \eqref{eq:finite energy state} and
\eqref{eq:basis Phi} one finds that the expectation value of $\calH_0$
for the state $\Phi$ is 
$
 E_{\Phi}  =  (\Phi,\calH_0\Phi)/(\Phi,\Phi) = -6s + 6sF\norm{\Phi}^{-2}, 
$
with
$
 {F = \sum_{I_\up,I_\dn\subset I\backslash\{\pi\}} 
             \left(|g(I_\up;I_\dn)|^2
                   -|\g{\pi}{\pi}|^2 \right)}
$
and 
${\norm{\Phi}^2=\left(\sum_{I_\up,I_\dn\subset I}
                         |g(I_\up;I_\dn)|^2
                  \right)}
$~\cite{comment:Phi},
where coefficients $g$ should satisfy
the condition~\eqref{eq:ccPhi}.
Here, and in what follows,  
we abbreviate  $I_{\sigma}\cup\{p\}$ as $I_{\sigma}^p$ 
for $p\in I$.
In the following,
we will show
$F\ge 0$.
This implies $E_\Phi\ge -6s$ since $s>0$.

To prove $F\ge0$, we first derive 
conditions on $g$ imposed by eq.~\eqref{eq:ccPhi}.
If we denote
$(\varphi_x^{(p)})^\ast=\{c_{x,\sigma},d_{p,\sigma}^\dagger\}$,
the left-hand-side of eq.~\eqref{eq:ccPhi} becomes
\begin{eqnarray}
\label{eq:ccPhi left}
\lefteqn{\sum_{I_\up,I_\dn\subset {I};|I_\up|\ge1,|I_\dn|\ge1}
          g(I_\up;I_\dn)
          \sum_{p\in I_\up}\sum_{p^\prime\in I_\dn}
	  (-1)^{|I_\up|-1}\mathsf{S}_{I_\up}^p\mathsf{S}_{I_\dn}^{p^\prime}}\nonumber\\
&&    \hspace*{7em}\times (\varphi_x^{(p)})^\ast(\varphi_x^{(p^\prime)})^\ast
          \Phi(I_\up\backslash\{p\};I_\dn\backslash\{p^\prime\})\nonumber\\ 
&=&
 \sum_{p,p^\prime\in {I}}
  (\varphi_x^{(p)})^\ast(\varphi_x^{(p^\prime)})^\ast \nonumber\\
&&
  \times\sum_{I_\up\subset {I}\backslash\{p\}}
  \sum_{I_\dn\subset {I}\backslash\{p^\prime\}}
          (-1)^{|I_\up|}
	  \mathsf{S}_{I_\up}^p\mathsf{S}_{I_\dn}^{p^\prime}	  
          g(I_\up^p;I_\dn^{p^\prime})          
          \Phi(I_\up;I_\dn) \nonumber\\
 &=&\sum_{I_\up,I_\dn\subset {I}}
 \sum_{p,p^\prime\in {I}}
  (\varphi_x^{(p)})^\ast(\varphi_x^{(p^\prime)})^\ast 
          \tg{p}{p^\prime}\Phi(I_\up;I_\dn)\,, 
\end{eqnarray}
where 
$\mathsf{S}_{I_\sigma}^p$, 
which corresponds to a sign factor coming {from}
exchange of the fermion operators,
equals 1 if $\sum_{p^\prime\in I_\sigma;p^\prime<p}1$ is
even and $-1$ otherwise. 
In the final expression of \eqref{eq:ccPhi left}, we introduced  
subsidiary coefficients $\tilde{g}$ defined as 
$\tg{p}{p^\prime}=0$ if $p\in I_\up$ or $p^\prime\in I_\dn$ and
$
 \tg{p}{p^\prime} =(-1)^{|I_\up|}
			   \mathsf{S}_{I_\up}^p\mathsf{S}_{I_\dn}^{p^\prime}
			   \g{p}{p^\prime}
$ 
otherwise.
Therefore, 
$c_{x,\dn}c_{x,\up}\Phi=0$ holds if and only if
$
\sum_{p,p^\prime\in {I}}
  (\varphi_x^{(p)})^\ast(\varphi_x^{(p^\prime)})^\ast 
          \tg{p}{p^\prime} = 0
$
for any ${I_\up,I_\dn\subset I}$.
Taking the sum of this equation over $x\in C_0^{1}$, 
we find that
$
 \sum_{p\in {I}} \frac{1}{(3+2\cos p)} \tg{p}{-p} = 0
$
and similarly taking the sum
over $x\in C_0^{2}$, we find that
$
\sum_{p\in {I}} \frac{(1+\cos p)}{(3+2\cos p)} \tg{p}{-p} = 0 
$
(where we identified $-\pi$ with $\pi$).
By eliminating $\tg{0}{0}$ 
{from} these two equations,
we obtain
\begin{eqnarray}
\tg{\pi}{\pi} &=& 
   -\frac{1}{16}\tg{\frac{\pi}{3}}{-\frac{\pi}{3}} 
   -\frac{1}{16}\tg{-\frac{\pi}{3}}{\frac{\pi}{3}}
\nonumber\\
\label{eq:final condition}
  && -\frac{3}{8}\tg{\frac{2\pi}{3}}{-\frac{2\pi}{3}} 
   -\frac{3}{8}\tg{-\frac{2\pi}{3}}{\frac{2\pi}{3}}\,.   
\end{eqnarray}
Our analysis below  relies heavily on this condition.

For a subset $I_\dn$ of ${I}$, we define
$
 \bar{I}_{\dn} = \{-p~|~p\in I_\dn \}
$
and denote by $N(I_\up;I_\dn)$ the number of elements in 
${I_\up\cap \bar{I}_\dn \cap ({I}\backslash\{0,\pi\})}$.
Condition~\eqref{eq:final condition} relates 
$\tilde{g}(I_\up;I_\dn)$
with $I_\up,I_\dn$ such that $N(I_\up;I_\dn)=r$ 
and $\tilde{g}(I_\up^\prime;I_\dn^\prime)$
with $I_\up^\prime,I_\dn^\prime$ such that 
$N(I_\up^\prime;I_\dn^\prime)=r+1$.
This motivates us to decompose $F$ as
$F= F^\prime+\sum_{r=0}^{4} F_r$, where
\begin{eqnarray}
 F_r&=&\sumtwo{I_\up,I_\dn\subset {I}\backslash\{\pi\};}
                  {N(I_\up;I_\dn)=r+1}
                  |g(I_\up;I_\dn)|^2 
          -\sumtwo{I_\up,I_\dn\subset {I}\backslash\{\pi\};}
                  {N(I_\up;I_\dn)=r}
                           |\g{\pi}{\pi}|^2 
                                         \,,\nonumber\vspace*{-2\baselineskip}\\
                                                    \\
 F^\prime &=& \sum_{I_\up,I_\dn\subset {I}\backslash\{\pi\};
                         N(I_\up;I_\dn)=0}|g(I_\up;I_\dn)|^2\,.
\end{eqnarray}
Since the term $F^\prime$ is apparently non-negative, 
$F\ge0$ is implied by $F_r\ge0$ for $r=0,\dots,4$.

We shall prove that $F_r\ge0$ by using~\eqref{eq:final condition}.
For a pair of $I_\up^{\pi}$ and $I_\dn^{\pi}$ such that 
$N(I_\up^\pi;I_\dn^\pi)=r$, 
the number of non-zero $\tilde{g}$ in the
right hand side of \eqref{eq:final condition} is, by the definition,
at most $4-r$, and thus for such a pair we have~\cite{comment:inequality}
\begin{equation}
\label{eq:final condition inequality}
|\tg{\pi}{\pi}|^2  
          \le  \frac{9}{64}(4-r)\sum_{p\in {I}\backslash\{0,\pi\}}
                    |\tg{p}{-p}|^2\,. 
\end{equation}
Then, we find that
\begin{eqnarray}
\lefteqn{\sumtwo{I_\up,I_\dn\subset {I}\backslash\{\pi\};}
                  {N(I_\up;I_\dn)=r}
                           |\g{\pi}{\pi}|^2
= \sumtwo{I_\up,I_\dn\subset {I}\backslash\{\pi\};}
                  {N(I_\up;I_\dn)=r}
                           |\tg{\pi}{\pi}|^2} \nonumber\\
&\le&
\frac{9}{64}(4-r)\sumtwo{I_\up,I_\dn\subset {I}\backslash\{\pi\};}
                  {N(I_\up;I_\dn)=r}\sum_{p\in {I}\backslash\{0,\pi\}}
                    |\tg{p}{-p}|^2 \nonumber\\
&=&\frac{9}{64}(4-r)(r+1)\sumtwo{I_\up,I_\dn\subset {I}\backslash\{\pi\};}
                  {N(I_\up;I_\dn)=r+1}
                           |g(I_\up;I_\dn)|^2 \nonumber\\
&\le& \frac{27}{32}\sum_{I_\up,I_\dn\subset {I}\backslash\{\pi\};
                              N(I_\up;I_\dn)=r+1}
                           |g(I_\up;I_\dn)|^2\,.
\label{eq:tg inequality}
\end{eqnarray}
To get the third line, we have used the fact that,
for $I_\up$ and $I_\dn$ such that $N(I_\up;I_\dn)=r+1$,
there are $r+1$ elements $p$ in ${I}\backslash\{0,\pi\}$ for which
we can find suitable subsets $I_\up^\prime$ and $I_\dn^\prime$ 
such that ${\{p\}\cup I_\up^\prime=I_\up}$ and
${\{-p\}\cup I_\dn^\prime=I_\dn}$.   
To obtain the final inequality, we have used 
$(4-r)(r+1)\le6$  for $0\le r \le 4$.
By using~\eqref{eq:tg inequality} we obtain
\begin{equation}
 F_r \ge \frac{5}{32}\sum_{I_\up,I_\dn\subset {I}\backslash\{\pi\};
                  N(I_\up;I_\dn)=r+1}
                           |g(I_\up;I_\dn)|^2\ge0\,.
\end{equation}

We therefore conclude that $F\ge 0$. 
The above analysis also shows that the equality $F=0$
holds only when $F^\prime$ and $F_r$ are
vanishing, i.e., $g(I_\up;I_\dn)=0$ for any pair of $I_\up$ and $I_\dn$  
such that $\pi \in I_\up\cap I_\dn$ or 
$\pi \notin I_\up\cup I_\dn$.

In other words we have shown that $E_\Phi\ge -6s$
for any $\Phi$ and 
that any $\Phi$ attaining the minimum expectation value $-6s$ is written as  
\begin{equation}
\label{eq:Phi}
 \Phi = \sum_{I_\up,I_\dn\subset {I};
               \pi\in I_\up\cup I_\dn,\pi\notin I_\up\cap I_\dn}
        g(I_\up;I_\dn)\Phi(I_\up;I_\dn)\,
\end{equation}
and further satisfies the finite energy  
condition~\eqref{eq:ccPhi}.
One finds that such minimizing $\Phi$ indeed exists by testing
$d_{\pi,\up}^\dagger{\Phi}_0$ or 
$\prod_{p\in I}d_{p,\up}^\dagger{\Phi}_0$.
By construction such $\Phi$ is an eigenstate of
$\calH_0$ as well as 
${-s\sum_{\sigma}a_{0,\sigma}^\dagger a_{0,\sigma}}$.
Since it is known to be the lowest energy state of $\calH_0$
in the limit $t,U\to\infty$, the continuity of energy implies
that such  
$\Phi$ is the lowest energy state of $\calH_0$ for sufficiently large 
$t/s$ and $U/s$.
It is also easy to check  
that such $\Phi$ has 
the properties stated in Lemma.
(Note that $d_{\pi,\sigma}=\frac{\mu[0,0]}{\sqrt{6}}a_{0,\sigma}$.)
This completes the proof of Lemma. \rule{5pt}{5pt}        

\textit{Proof of Theorem}.
We assume that the values of $t/s$ and $U/s$ are
large enough for the statement in Lemma to hold.
We note that how large $t/s$ and $U/s$ should be is independent of
the size of $\La$, because Lemma is 
concerned with
the local
Hamiltonian.

{From} Lemma we find that 
the eigenvalue of $\calH$ is bounded below by $-6s|\calL|$,
while, by taking 
${\Phi_\mathrm{f}=\prod_{\alpha\in\calL}a_{\alpha,\up}^\dagger\Phi_0}$
as a variational state, 
we find that $-6s|\calL|$ is
an upper bound on the ground state energy.
Therefore,
the ground state energy is $-6s|\calL|$, and 
$\Phi_\mathrm{f}$ and its SU(2) rotations 
are among the corresponding eigenstates.
It is 
apparent that these states have $\Stot=|\calL|/2$.

The remaining task is to prove the uniqueness.   
Let $\PhiG$ be an arbitrary ground state of $\calH$.
Lemma implies that the ground state energy is attained if and
only if ${\calH_\alpha\PhiG=-6s\PhiG}$ for all $\alpha\in\calL$.
Thus $\PhiG$ must satisfy 
the conditions stated in Lemma.

The condition \eqref{eq:condition1 in Lemma} implies that $\PhiG$
is expressed as~\cite{comment:basis} 
\begin{equation}
\label{eq:ground state}
 \PhiG = \sum_{\{\sigma\}} \varphi(\{\sigma\})
         \prod_{\alpha\in \calL}
           a_{\alpha,\sigma_\alpha}^\dagger \Phi_0 \,,
\end{equation}
where $\{\sigma\}$ is a shorthand for a spin configuration
$\{\sigma_\alpha\}_{\alpha\in \calL}$, the summation 
is over $\sigma_\alpha=\up,\dn$ for all $\alpha\in\calL$, 
and $\varphi(\{\sigma\})$
is a complex coefficient. 

Let us impose the condition~\eqref{eq:condition2 in Lemma} on $\PhiG$
in the form of~\eqref{eq:ground state}. 
Let $\beta$ and $\gamma$ be nearest neighbour points 
in $\calL$, i.e., $|\beta-\gamma|=1$,
and let $m(\beta,\gamma)\in\La$ 
be the site located at the mid-point between $\beta$
and $\gamma$.
It is easy to see that
$\{c_{m(\beta,\gamma),\sigma},a_{\alpha,\sigma}^\dagger\}$ is nonvanishing
if $\alpha=\beta$ or $\gamma$, and is vanishing otherwise.
Then, it follows {from} 
the condition 
$c_{m(\beta,\gamma),\dn}c_{m(\beta,\gamma),\up}\PhiG=0$ 
that $\varphi(\{\sigma\})
=\varphi(\{\tau\})$
for any pair of spin configurations $\{\sigma\}$ and $\{\tau\}$
satisfying that $\sigma_\beta=\tau_\gamma$, $\sigma_\gamma=\tau_\beta$,
and $\sigma_\alpha=\tau_\alpha$ for $\alpha\ne\beta,\gamma$.
Examining $c_{m(\beta,\gamma),\dn}c_{m(\beta,\gamma),\up}\PhiG=0$ for
all the pairs of nearest neighbour points in $\calL$, we find
that $\varphi(\{\sigma\})=\varphi(\{\tau\})$ 
whenever $\sum_{\alpha}\sigma_\alpha=\sum_{\alpha}\tau_\alpha$. 
Therefore 
$\PhiG$ is written as
$
 \PhiG=\sum_{M=0}^{|\calL|}\varphi_M
 (S_\mathrm{tot}^-)^M\Phi_\mathrm{f}\,,
$
where $\varphi_M$ are new coefficients and 
the spin lowering operator $S_\mathrm{tot}^-$ is defined by 
$S_\mathrm{tot}^-=\sum_{x\in\La} c_{x,\dn}^\dagger c_{x,\up}$. 
This completes the proof.~\rule{5pt}{5pt}

We would like to thank Professor T. Idogaki for continuous encouragement.
We would like to thank H. Tasaki and T. Sekizawa for stimulating
discussions and for sending us their papers prior to publication.
We also thank H. Tasaki for many helpful comments on the manuscript.



\begin{thebibliography}{99}
\bibitem{comment:other-line-graphs}
	It is possible to prove stability of ferromagnetism in 
	Mielke's models on other lattices such as
	a regular lattice of corner-sharing tetrahedra.
	The details will appear elsewhere.
\bibitem{comment:norm}
	Here $|\cdot|$ denotes the Euclidean norm. 
        We use the same symbol $|X|$ to denote the number of elements in a set $X$.
\bibitem{comment:Phi}
	Note that $\Phi(I_\up;I_\dn)$ are the eigenstates of 
	$-s\sum_\sigma a_{0,\sigma}^\dagger a_{0,\sigma}$
	and  the set of these states is orthonormal.
\bibitem{comment:inequality}
	We have used inequality  
	$\left|\sum_{i=1}^N z_i \right|^2
	\le N \sum_{i=1}^N |z_i|^2$
	which follows {from} the Schwarz inequality.
\bibitem{comment:basis}
	Proof: Since all $a_{\alpha,\sigma}^\dagger\Phi_0$ with $\alpha\in\calL$
	are linearly independent, we can form a basis of 
        the single-electron Hilbert space on $\La$ by adding  
	$|\La|-|\calL|$ linearly independent states. 
	We denote these states by $f_{i,\sigma}^\dagger\Phi_0$
	with $i\in\calI$ where $\calI$ is some index set with $|\calI|=|\La|-|\calL|$.
	Then, the set of states 
	{\footnotesize
	$
	  \left(\prod_{\alpha\in\calL_\up}a_{\alpha,\up}^\dagger\right)
	  \left(\prod_{\beta\in\calL_\dn}a_{\beta,\dn}^\dagger\right)
	  \left(\prod_{i\in\calI_\up}f_{i,\up}^\dagger\right)
	  \left(\prod_{j\in\calI_\dn}f_{j,\dn}^\dagger\right)\Phi_0
	$}
	with $\calL_\up,\calL_\dn\subset \calL$ and $\calI_\up,\calI_\dn\subset \calI$
	satisfying 
	$\sum_{\sigma}(|\calL_\sigma|+|\calI_\sigma|)=\Ne$ is a
	basis of 
        $\Ne$-electron Hilbert space. 
	We suppose that $\PhiG$ is
	expanded with respect to this basis and then use
	condition~\eqref{eq:condition1 in Lemma}.
\end{thebibliography}
\end{document}